\documentclass[conference]{IEEEtran}
\IEEEoverridecommandlockouts
\usepackage{cite}
\usepackage{amsmath,amssymb,amsfonts}
\usepackage{amsthm}
\usepackage{algorithmic}
\usepackage{graphicx}
\usepackage{textcomp}
\usepackage{xcolor}
\usepackage{makecell}
\usepackage{bm}
\usepackage{subfigure}

\def\BibTeX{{\rm B\kern-.05em{\sc i\kern-.025em b}\kern-.08em
    T\kern-.1667em\lower.7ex\hbox{E}\kern-.125emX}}
\begin{document}

\title{Communication and Computation Assisted Sensing Information Freshness Performance Analysis in Vehicular Networks\\}
\author{\IEEEauthorblockN{Ning Jiang, Shi Yan, Zhuohan Liu, Chunjing Hu, Mugen Peng }
\IEEEauthorblockA{State Key Laboratory of Networking and Switching Technology, \\
		Beijing University of Posts and Telecommunications, Beijing, 100876, China. \\ 
		Email:$\{$creativejn,yanshi01,liuzhuohan,hucj,pmg$\}$@bupt.edu.cn\\}} 
\maketitle

\begin{abstract}
The timely sharing of raw sensing information in the vehicular networks (VNETs) is essential to safety. In order to improve the freshness of sensing information, joint scheduling of multi-dimensional resources such as communication and computation is required. However, the complex relevance among multi-dimensional resources is still unclear, and it is difficult to achieve efficient resource utilization. In this paper, we present a theoretical analysis for a novel metric Age of Information (AoI) on a communication and computation assisted spatial-temporal model. An uplink VNETs scenario where Road Side Units (RSUs) are deployed with computational resource is considered. The transmission and computation process is unified into a two-stage tandem queue and the expression of the average AoI is derived. The network interference is analyzed by modeling the VNETs as Cox Poisson Point Process based on stochastic geometry and the closed-form solution of the coverage probability and the expected data rate performance under the constraints of transmission resources is obtained. The simulation results reveal the basic relationship between communication and computation capacity and show that communication and computation should reach a tradeoff to improve resource utilization while ensuring real-time information requirement.
\end{abstract}


\section{Introduction}

The vehicular networks (VNETs) have been identified as one of the  typical application scenarios of intelligent transportation system in the future which can support a variety of emerging services \cite{1}[2]. For time-sensitive applications, vehicles need to continuously share the  environment status and road safety traffic to the road side unit (RSU) for analyzing the collected data to obtain real-time sensing information, thereby assisting the intelligent coordinated control. Age of Information (AoI) is a new key performance indicator (KPI) recently proposed for characterizing the freshness of the data \cite{3}. In order to make adequate deployment, it is not only necessary to understand how the network affects the timeliness of information transmission, but also to extract the status information embedded in the data packet.

Recently, AoI has been used in communication systems for various real-time applications (including VNETs) due to the high sensitivity to data freshness. The expression of peak and average AoI with preemption was deduced in \cite{4}, where the influence of channel access control (such as ALOHA) was considered. In \cite{5}, the authors derived the peak AoI of large-scale IoT uplink networks under time-triggered and event-triggered traffic. In \cite{6}, the authors presented the AoI of computation intensive data in edge computing, and considered user local computing and remote computing on edge servers. 

Although some scholars have carried out some research on AoI in wireless networks, the related research has the following limitations: Although there are many analysis models based on communication process, it may not be suitable to describe the actual traffic processing of data packets because the computation also impacts on the information freshness performance. Moreover, the freshness of information in large-scale wireless networks is usually leveraged based on queuing, while queuing analysis only captures time influence between transmitters and receivers \cite{7}. Due to interference, the AoI is related with spatial distribution of active transmitter nodes, leading to the queue of the link to influence each other in space and time. In this paper, we investigate the communication and computation assisted spatial-temporal AoI performance in VNETs. The main contributions are as follows:
\begin{itemize}	
\item  A VNETs model is considered to establish spatial-temporal model using stochastic geometry and queuing theory to represent the macro and micro network performance: the intelligent vehicles transmit sensing data sampled by on-board sensors and implement data demodulation process in RSUs \cite{8}.	
\item A tandem queue model is proposed for the two stages of transmission and computation, and the analytical expression of the average AoI is derived. The Cox point process is used to represent the distribution of vehicles and RSUs \cite{9}. Some closed-form solutions of KPIs such as uplink coverage probability and expected data rate are obtained.	
\item The simulation results evaluate the accuracy of the analysis and obtain the tradeoff between transmission and computation.
\end{itemize} 

\section{SYSTEM MODEL}
\subsection{Vehicular Networks Spatial Model}

\begin{figure}[!htp]
	\vspace*{-1em} 
	\centering  
	\setlength{\abovecaptionskip}{0.cm}
	\subfigure[Spatial-temporal VNETs interference model.]{   
		\begin{minipage}{4cm}
			\centering    
			\includegraphics[scale=0.4]{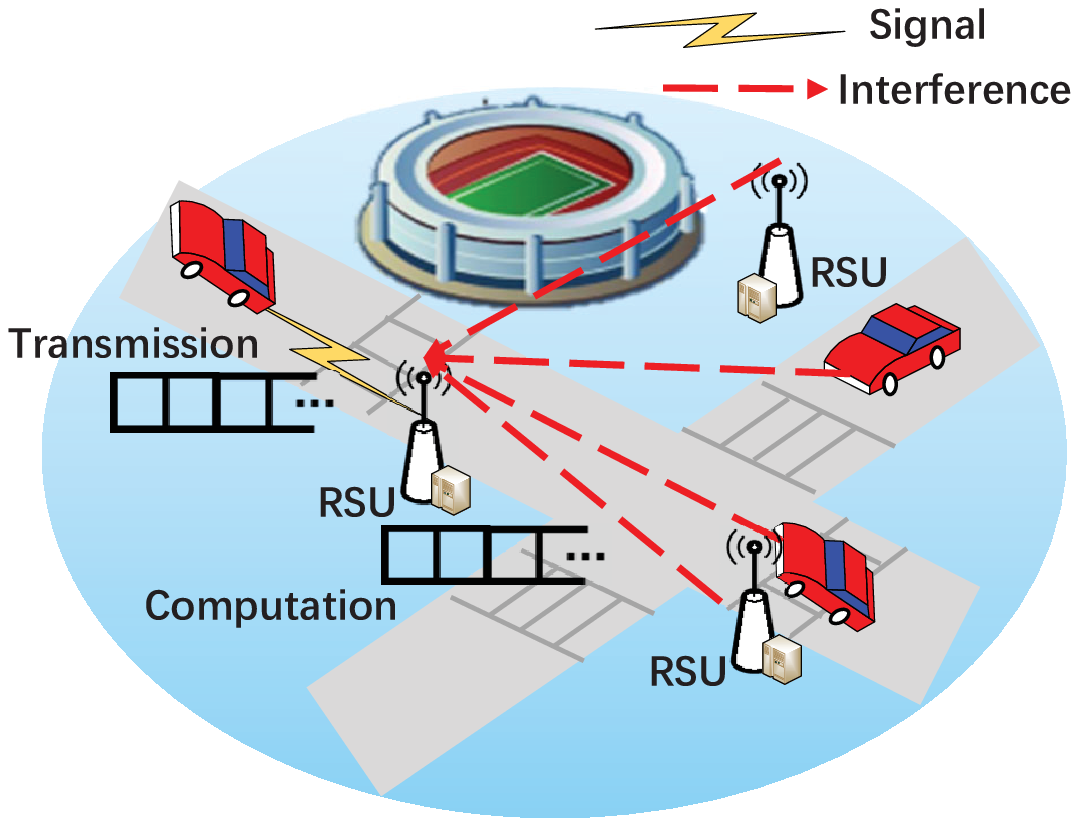}  
		\end{minipage}
	}
	\subfigure[Illustration of the Cox PPP of vehicles and RSUs.]{ 
		\begin{minipage}{4cm}
			\centering    
			\includegraphics[scale=0.4]{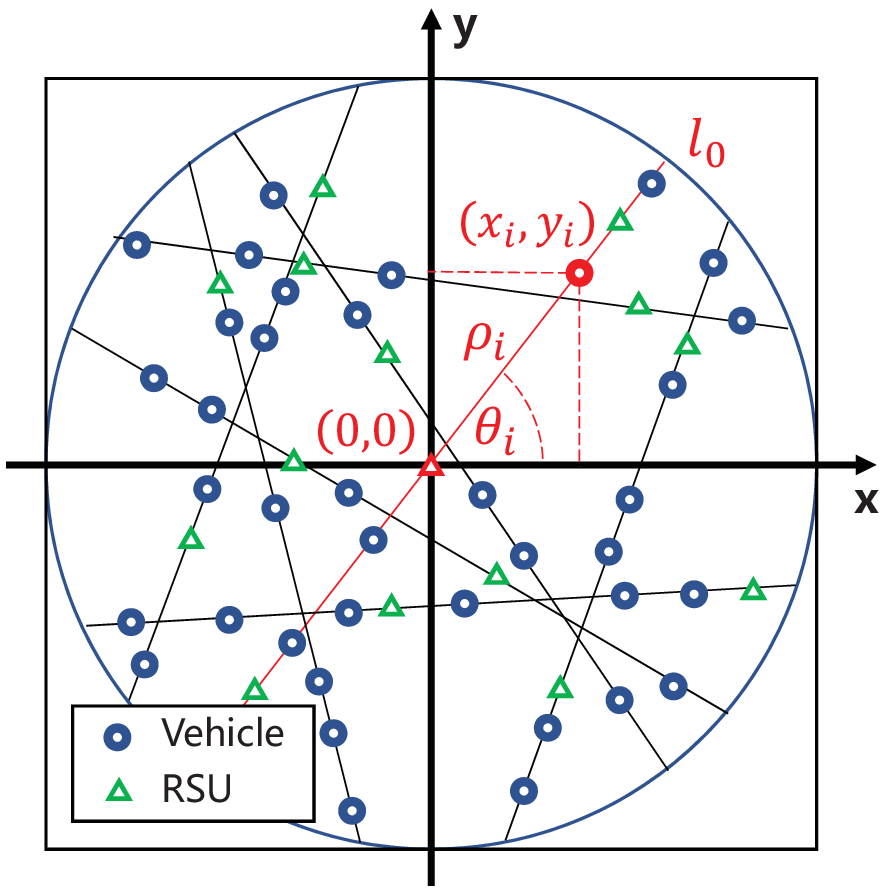}
		\end{minipage}
	}
	\caption{Spatial-temporal system model of the uplink VNETs based on queue and Cox PPP.}   
	\vspace*{-0.1cm} 
	\label{fig1}    
\end{figure}

As shown in Fig.~\ref{fig1}(a), an uplink cellular-based VNETs model is considered. The intelligent vehicles can acquire sensing information and transmit status update traffic to its associated RSU which is deployed with computational resource. We model the vehicles and RSUs by the Cox Poisson Point Process (PPP) which is illustrated in Fig.~\ref{fig1}(b). The road is modeled as an independent motion-invariant Poisson line process $\phi _l$ with line intensity $\mu _l$, which can be represented on a cylindrical space $\mathbf{C}:=\mathbb{R} \times (0,\pi )$ generated by a homogeneous point process $\varPsi$ with intensity $\lambda _l=\mu _l / {\pi}$. Each node $(\rho _i,\theta _i)$ in $\varPsi$ is associated with the corresponding line $L_i$ as follows
\begin{equation}
	L_i\left( \rho _i,\theta _i \right) \! =\! \left\{ \left( x,y \right) \in \mathbb{R} ^2 \!\mid\! xcos\left( \theta _i \right) \! + \!ysin\left( \theta _i \right)  \!= \!\rho _i \right\} ,
	\label{eq1}
\end{equation}
where $\theta _i$ denotes the angle between the line $L_i$ and the positive direction of the $x$-axis, and $\rho _i$ denotes the distance between the line $L_i$ and the origin \cite{10}. By applying the Slivnyak theorem, the translation of the origin can be regarded as the process of adding a point to the space $\mathbf{C}$\cite{11}. The tagged receive node is located at the origin. Therefore, we can set $l_0$ to be the line containing the tagged nodes, which represents the tagged line, and $\phi_{l_0}$ be the 1-D Poisson point process on this line. $\varPhi$ is set to represent the Cox PPP which contains the all transmission nodes in VNETs.

Next, the vehicles on each line are modeled as an independent one-dimensional (1-D) PPP with intensity $\lambda _v$ and the RSUs are modeled with intensity $\lambda _r$ by a similar process. Without loss of generality, we assume that the ratio of nodes in the transmitting state to all nodes is $p$ and the transmission vehicles work on the same frequency band as the RSUs. The distribution of the nodes on the line is a PPP with intensity $\lambda _p=\lambda _{pv}+\lambda _{pr}$\cite{8}. The vehicle transmission power is $P _v$ and only one antenna is configured. 

The interference of the wireless channel can be expressed by a combination of standard path loss and small-scale fast fading, where the standard path loss can be expressed as $\left\| X_{r} \right\| ^{-\alpha}$, where $\left\| X_{r} \right\|$ represents the distance between a tagged transmitter and its related receiver. Considering the high proportion of the Line of Sight (LoS) component between the tagged nodes in the case of accessing the nearest RSU, the small-scale fast fading can be assumed to be Rician fading. The probability density function (PDF) of Rician fading can be given by \cite{12}
\begin{equation}
	\begin{array}{c}
		f_x\left( x \right) =\left( 1+K \right) e^{-K-\left( 1+K \right) x}I_0\left( 2\sqrt{K\left( 1+K \right)}x \right)\\
		\approx \sum_{i=1}^{N_K}{w_{i}^{K}e^{-u_{i}^{K}x}},x\in \left[ 0,W \right],\\
	\end{array}
	\label{eq2}
\end{equation}
where $I_0\left( \cdot \right)$ denotes zero-order modified Bessel function of the first kind, and $K$ denotes the Rician factor which represents the ratio of the power of the LoS component to the power of the diffuse component. $N_K$ is the number of weighted terms which is related to the Rician factor. Besides,the formula must conform to the two constraints  $\sum\nolimits_{i=1}^{N_K}{w_{i}^{K}}=1$ and $u_{i}^{K}>0$.

To analyze the wireless link, the expression of signal-to-interference ratio of the transmission link is given by
\begin{equation}
	\gamma _r=\frac{P_vh_{r}\left\| X_{r} \right\| ^{-\alpha}}{I_{{\phi_{l_0}}}+I_{\varPhi /{\phi_{l_0}}}},
\end{equation}
where $h_{r}$ and $\left\| X_{r} \right\| ^{-\alpha}$ respectively denote the Rician fading and path loss between the tagged transmitter and receiver. $I_{\phi_{l_0}}$ denotes interference from nodes on the same road and $I_{\varPhi /{\phi_{l_0}}}$ denotes the interference from nodes on different roads.
\subsection{Transmission and Computation Assisted Temporal Model}
\begin{figure}[htbp]
	\vspace*{-1em} 
	\centering
	\setlength{\abovecaptionskip}{0.cm}
	\includegraphics[width=0.35\textwidth]{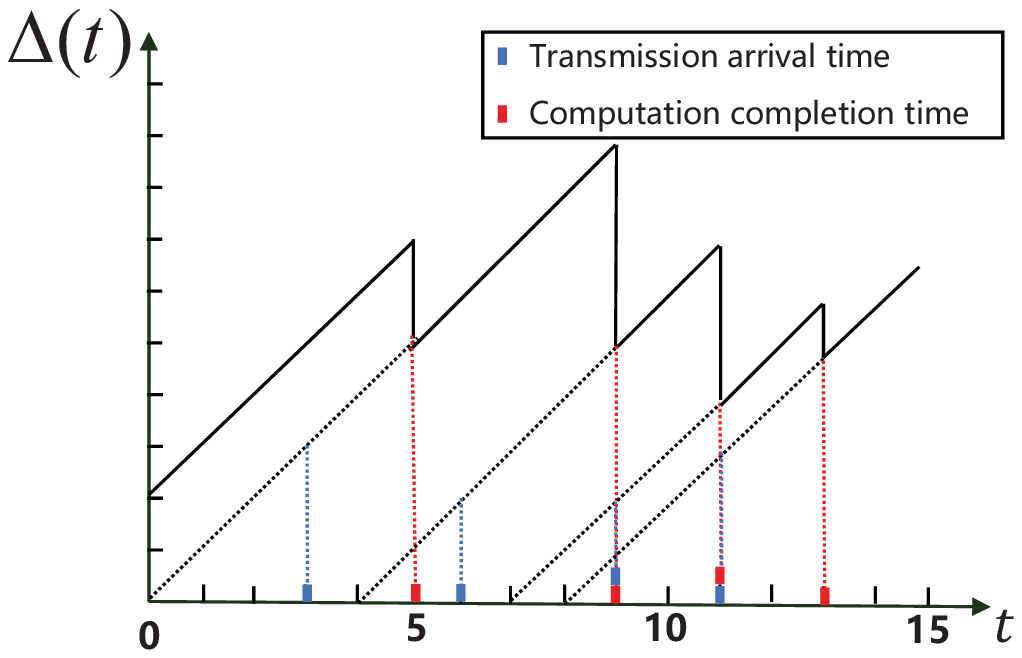}
	\caption{Age of Information for transmission and computation tandem queue.}
	\vspace*{-0.1cm}
	\label{fig2}
\end{figure}

A tagged link is formed as the transmitter-receiver pair. As depicted in Fig.~\ref{fig2}, AoI is employed to measure the performance of the transmission and computation assisted system which is used to quantify the information freshness of the process from the generation of sensing information to the end of computation. When RSU does not demodulate the content of the data packet, the AoI on the tagged link increases linearly with a slope of $1$. The AoI will reduce the time elapsed from the generation of the data packet to the completion of the computation. The process of data packages is subject to the First-Come-First-Served (FCFS) principle. The definition of AoI\cite{3} is $\Delta \left( t \right) =t-A\left( t \right)$, where $\Delta \left( t \right)$ denotes the AoI of the link, and $A\left( t \right)$ denotes the end time of the tandem process. Average AoI is usually used as a performance metric to evaluate the freshness of information which can be obtained as $avgAoI=\underset{T\rightarrow \infty}{\lim}\frac{1}{T}\int_0^T{\Delta}\left( t \right) \mathrm{d}t$\cite{3}.

The two-stage tandem M/M/1 queue is used to establish the transmission and computation assisted model. $
\mu _1=\frac{bR}{L}$ is the transmission rate where $b$ is the bandwidth, $L$ is the packet size in bit and $R$ is the expected data rate. The computation process is initiated after the transmission process ends. In the process unit of RSU, the CPU cycle frequency can be adjusted by voltage through dynamic voltage and frequency scaling (DVFS) technology, so the computation rate is $\mu _2=f/L\kappa$ and the computation power is $
P_c=\zeta f^3$ where $\kappa$ denotes the number of CPU cycles required to process one bit and $\zeta$ is the conversion factor depends on the average switched capacitance and the average activity\cite{13}.  

\section{PERFORMANCE ANALYSIS}
 In this section, we first obtain the transmission and computation  assisted AoI analytical expressions represented by the M/M/1 tandem queue. Then we derive the closed-form solution of coverage probability and the expected data rate representing the uplink transmission performance of the VNETs.

\subsection{Average Age of Information the Transmission-Computation tandem queue}\label{AA}
\begin{figure}[htbp]
	\vspace*{-1em}
	\centering
	\includegraphics[width=0.48\textwidth]{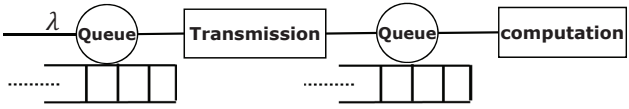}
	\caption{The transmission-computation tandem queue for vehicle and tagged RSU.}
	\vspace*{-1em}
	\label{fig3}
\end{figure}

As shown in Fig.~\ref{fig3}, the information traffic from the source is transmitted in turn in the first queue, and then reaches the second queue to continue computation process. It is assumed that the traffic arriving at the first queue follows Poisson distribution, which means the sampling time interval of different data packets is the negative exponential distribution with the parameter $\lambda$.

The M/M/1 tandem queue conforms to the overtake-free and quasi-reversible properties\cite{14}. An important feature of the quasi-reversible queue is that the Poisson traffic passing through the queue is statistically invariant and can produce a Poisson output traffic at the same rate, in other words, Poisson-in-Poisson-out. Therefore, the interaction between the transmission queue and the computation queue can be ignored when calculating the average AoI, and an approximate tractable result can be obtained.

\begin{figure}[htbp]
	\setlength{\abovecaptionskip}{0.cm}
	\centering
	\includegraphics[width=0.48\textwidth]{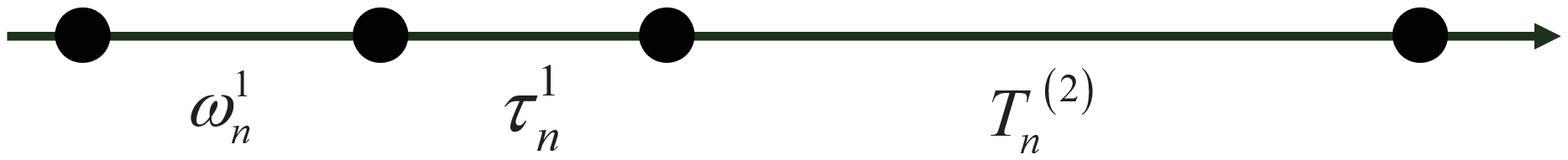}
	\caption{Packet process flow in sensing information traffic.}
	\label{fig4}
\end{figure}

Next, for a certain data packet in the system, the process is shown in Fig.~\ref{fig4}. $\omega _{n}^{1}$ denotes the queuing time of the first queue, $\tau _{n}^{1}$ denotes the transmission time of the first queue and $T_{n}^{\left( 2 \right)}$ is the time of the packet in the second queue.
\begin{figure}[htbp]
	\setlength{\abovecaptionskip}{0.cm}
	\vspace*{-1em}
	\centering
	\includegraphics[width=0.45\textwidth]{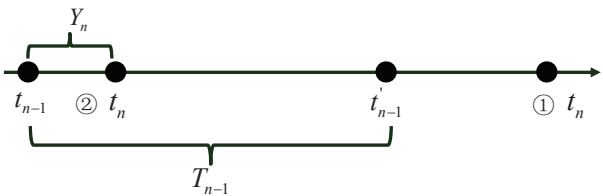}
	\caption{Time period division between adjacent packets.}
	\label{fig5}
\end{figure}
 
When calculating the average AoI, we only pay attention to the time interval distribution of each data packet sampling $Y_n$ and the system time distribution $T_n$ due to the fundamental formula $avgAoI={\left( E\left[ T_nY_n \right] +E\left[ \frac{Y_{n}^{2}}{2} \right] \right)}/{E\left[ Y_n \right]}$ in\cite{3}.

Therefore, the sum of the service time in the first queue and the system time in the second queue can be approximated as the overall service time $\tau _n$. Thus, $T_n=\omega _{n}^{1}+\tau _n$, in which $\tau _n$ and $T_n$ are independent. As shown in Fig.~\ref{fig5}, $\omega _{n}^{1}$ can be divided into two situations: if the current $n-1$ packets have been processed, $\omega _{n}^{1}$ is 0, and if the current $n-1$ packets have not been processed, $\omega _{n}^{1}=T_{n-1}-Y_n$. Thus the expectation of $\omega _{n}^{1}$ under condition $Y_n=y$ can be derived as
\begin{equation}
	\begin{array}{l}
		E\left[ \omega _{n}^{1}|Y_n=y \right] =E\left[ \left( T_{n-1}-Y_n \right) ^+|Y_n=y \right]\\
		=E\left[ \left( T-Y \right) ^+ \right] =\int_{y}^{\infty}{\left( t-y \right)}f_T\left( t \right) dt.\\
	\end{array}
	\label{eq13}
\end{equation}

Then,  the transmission rate and the computation rate are set to $\mu _1$ and $\mu _2$, respectively. Based on the statistical independence of the queue, when $\mu _1\ne \mu _2$, the distribution of the sojourn time of the tandem queue is given as
\begin{equation}
	\begin{array}{l}
		f_T(t)=\int_0^t{\left( \mu _1-\lambda \right)}e^{-\left( \mu _1-\lambda \right) \left( t-y \right)}\left( \mu _2-\lambda \right) e^{-\left( \mu _2-\lambda \right) y}dy\vspace{1ex} \\
		=\frac{\left( \mu _1-\lambda \right) \left( \mu _2-\lambda \right)}{\mu _1-\mu _2}\left[ e^{-\left( \mu _2-\lambda \right) t}-e^{-\left( \mu _1-\lambda \right) t} \right].\\
	\end{array}
	\label{eq14}
\end{equation}

When $Y_n=y$, the conditional expectation of $\omega _{n}^{1}$ is given as
\begin{equation}
	\begin{array}{l}
		E\left[ \omega _{n}^{1}|Y_n=y \right]=\int_y^{+\infty}{\eta \left( t-y \right)}\left[ e^{-\left( \mu _2-\lambda \right) t}-e^{-\left( \mu _1-\lambda \right) t} \right] dt\vspace{1ex} \\
		=\eta \left[ \frac{e^{-\left( \mu _2-\lambda \right) y}}{\left( \mu _2-\lambda \right) ^2}-\frac{e^{-\left( \mu _1-\lambda \right) y}}{\left( \mu _1-\lambda \right) ^2} \right],\\
	\end{array}
	\label{eq15}
\end{equation}
where $\eta =\frac{\left( \mu _1-\lambda \right) \left( \mu _2-\lambda \right)}{\mu _1-\mu _2}$.

So the expectation of $\omega _{n}^{1}Y_n$ is given as
\begin{equation}
\begin{array}{l}
		E\left[ \omega _{n}^{1}Y_n \right] =\int_0^{\infty}{y\cdot E\left[ \omega _{n}^{1}|Y_n=y \right] \cdot f_Y\left( y \right) dy}\vspace{1ex} \\
		=\int_0^{\infty}{y\cdot \eta \left[ \frac{e^{-\left( \mu _2-\lambda \right) y}}{\left( \mu _2-\lambda \right) ^2}-\frac{e^{-\left( \mu _1-\lambda \right) y}}{\left( \mu _1-\lambda \right) ^2} \right] \cdot \lambda e^{-\lambda y}dy}\vspace{1ex} \\
		=\lambda \eta \left[ \frac{1}{\left( \mu _2-\lambda \right) ^2\mu _{2}^{2}}-\frac{1}{\left( \mu _1-\lambda \right) ^2\mu _{1}^{2}} \right].\\
\end{array}
\label{eq16}
\end{equation}

The total service time expectation is approximately the sum of the service time expectation of the two queues, and the average AoI of the tandem queue is derived as follows
\begin{equation}
	\begin{array}{l}
		avgAoI=\frac{E\left[ \omega _{n}^{1}Y_n \right] +E\left[ \tau _nY_n \right] +E\left[ \frac{Y_{n}^{2}}{2} \right]}{E\left[ Y_{n} \right]}\vspace{1ex} \\
		\approx \frac{\lambda \eta \left[ \frac{1}{\left( \mu _2-\lambda \right) ^2\mu _{2}^{2}}-\frac{1}{\left( \mu _1-\lambda \right) ^2\mu _{1}^{2}} \right] +\frac{1}{\lambda}\cdot \left( \frac{1}{\mu _1}+\frac{1}{\mu _2} \right) +\frac{1}{\lambda ^2}}{\frac{1}{\lambda}}\vspace{1ex} \\
		=\lambda ^2\eta \left[ \frac{1}{\left( \mu _2-\lambda \right) ^2\mu _{2}^{2}}-\frac{1}{\left( \mu _1-\lambda \right) ^2\mu _{1}^{2}} \right] +\frac{1}{\mu _1}+\frac{1}{\mu _2}+\frac{1}{\lambda}.\\
	\end{array}
	\label{eq17}
\end{equation}

The average AoI in the tandem queue is given as follows when $\mu _1=\mu _2$
\begin{equation}
	avgAoI\approx \frac{2\lambda ^2}{\mu ^3}+\frac{2\lambda ^2}{\left( \mu -\lambda \right) \mu ^2}+\frac{2}{\mu}+\frac{1}{\lambda}.
	\label{eq18}
\end{equation}
\begin{proof}
	See Appendix A.
\end{proof}

\subsection{Transmission Data Rate}
To facilitate the performance analysis of the proposed model, we first derive the uplink coverage probability $P\left( P_v,T \right)$ where $P_v$ is the vehicle transmission power and $T$ is the threshold. $T$ can be regarded as an indicator that reflects quality-of-service (QoS). The larger value of $T$ may represent higher requirement for SIR, which means that better received signal quality is required to correctly demodulate useful signals. The uplink coverage probability is given as
\begin{equation}
	\begin{aligned}	
	P\left( P_v,T \right) & =\sum_{i=1}^{N_K}{w_{i}^{K}}\exp \left( -\varepsilon _1\left( \lambda _{pr}\left( \frac{P_r}{P_v} \right) ^{\frac{1}{\alpha}}+\lambda _{pv} \right) \right. 
	\\
	&    \left. -\varepsilon _2\left( \lambda _{pr}\left( \frac{P_r}{P_v} \right) ^{\frac{2}{\alpha}}+\lambda _{pv} \right) \right), 
	\end{aligned}
\end{equation}
where $\varepsilon _1=\frac{2\pi\left( u_{i}^{K}T \right) ^{\frac{1}{\alpha}}\left. \left\| X_r \right\| \right. }{\sin \left( \frac{\pi}{\alpha} \right) \alpha}$ and $
\varepsilon _2=\frac{2\pi ^3\lambda _l\left( u_{i}^{K}T \right) ^{\frac{2}{\alpha}}\left\| X_r \right\| ^2}{\sin \left( \frac{2\pi}{\alpha} \right) \alpha}$.

When $T$ and $P_v$ is fixed, the closed-form solution of the expected data rate can be obtained according to $R=\log _2\left( 1+T \right) P\left( P_v,T \right)$ .

\begin{proof}
	See Appendix B.
\end{proof}

\section{NUMERICAL RESULTS}
In this section, we present the simulation results of the following three parts: the average AoI of the transmission-computation tandem queue; the coverage probability and expected data rate of the uplink VNETs; the average AoI with different transmission and computation capability. Monte Carlo method is used to verify the correctness of theoretical analysis. The simulation range is a circular area with radius $1$ km, the intensity of roads $\lambda _l$ is $5\times 10^{-3}/{\pi}$, the distance between the tagged nodes $\left\| X_r \right\|$ is $20$ m. The intensity of transmission vehicle $\lambda _{pv}$ and RSU $\lambda _{pr}$ are $1\times 10^{-2}$ and $3\times 10^{-3}$, respectively. The bandwidth $b$ is $2$ MHz. The data packet sampling rate $\lambda$ is $100$/s with packet size $L$ set to be $\rm 1\times 10^3$ bits. Path loss exponent $\alpha$ is set to $3$, and RSU transmission power $P_r$ is $33$ dBm. We assume that RSUs are deployed with sufficient computation resources which the CPU cycles required for computing one input bit $\kappa$ is $1\times 10^3$ cycles/bit \cite{15}. The considered communication resource is the vehicle transmission power $P_v$ in the range of $\left[ 15,30 \right]$ dBm and the considered computation resource is CPU cycle frequency $f$ which is $\left[ 0.15,0.45 \right] \times 10^9$ cycles/bit. The conversion factor $\zeta$ is $1.25\times 10^{-26}$ J/cycle. The Rician fading coefficients $w_{i}^{K}$ and $u_{i}^{K}$\cite{13} are listed in Table I.

\begin{figure}[htbp]
	\vspace{-0.3cm}
	\setlength{\abovecaptionskip}{0.cm}
	\centering
	\includegraphics[width=0.45\textwidth]{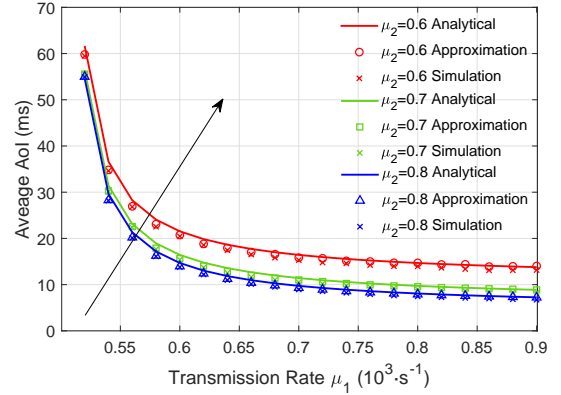}
	\caption{Average AoI with different transmission rates versus computation rate.}
	\vspace{-0.3cm}
	\label{fig6}
\end{figure}

\begin{table}[!htp]
	\vspace*{-1em}
	\renewcommand{\arraystretch}{1.3}
	\caption{RICIAN FADING SIMULATION PARAMETERS} \label{table_example}
	\centering
	\begin{tabular}{{|c|c|c|c|c|c|c|}}
		\hline
		Term Index& $n=1$ & $n=2$ & $n=3$ & $n=4$\\
		\hline
		$w_{i}^{K}$& -0.8993 & 5.9324 & -5.4477 & 1.4145\\
		\hline
		$u_{i}^{K}$& 1.2475 & 1.4298 & 1.7436 & 2.0326 \\
		\hline
	\end{tabular}\vspace*{-1em}
\end{table}
Fig.~\ref{fig6} shows the average AoI with different transmission rates and computation rates when the data packet sampling rate is fixed to $500$ /s for obvious comparison. The approximation is based on the statistical characteristics of the Poisson-in-Poisson-out of the tandem M/M/1 and the queue interaction is ignored. The analytical results are consistent with the results of the simulations which confirms the validity of the assumption and the accuracy of expressions. When the transmission and computation rate increase, the average AoI will decrease. Moreover, the corresponding law of the two rates to the average AoI is the same. In other words, increasing the same amount of transmission or computation rate will result the same reduction in average AoI. Further analysis shows that with the continuous increase of transmission rate and computation rate, the reduction degree of the average AoI becomes smaller and tends to be stable.

\begin{figure}[htbp]
	\vspace{-0.3cm}
	\setlength{\abovecaptionskip}{0.cm}
	\centering
	\includegraphics[width=0.45\textwidth]{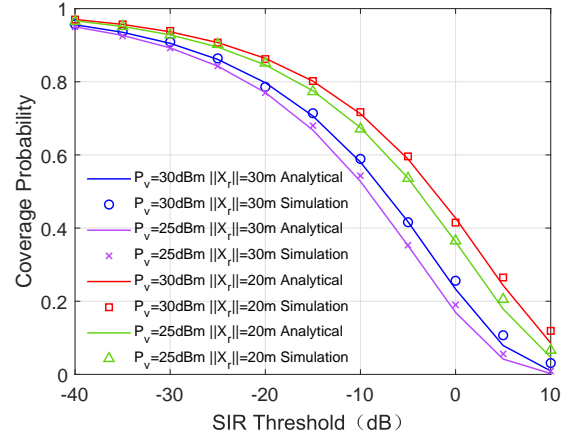}
	\caption{Coverage probability with different SIR threshold versus vehicle transmission power and $\left\| X_r \right\|$.}
	\vspace{-0.3cm}
	\label{fig7}
\end{figure}

\begin{figure}[htbp]
	\vspace{-0.3cm}
	\setlength{\abovecaptionskip}{0.cm}
	\centering
	\includegraphics[width=0.45\textwidth]{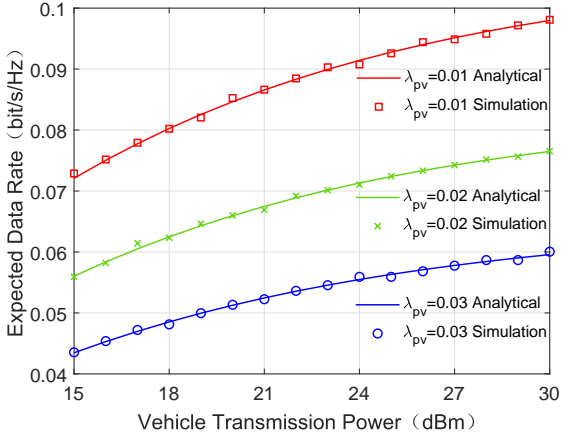}
	\caption{Expected data rate with different vehicle transmission power versus intensity of transmit vehicle.}
	\vspace{-0.3cm}
	\label{fig8}
\end{figure}

Fig.~\ref{fig7} and Fig.~\ref{fig8} show the uplink coverage probability and the expected data rate. As can be seen in Fig.~\ref{fig7}, for any vehicle transmission power and the distance between the tagged transmitter-receiver, when the SIR threshold increases, the coverage probability decreases. For a larger distance between a vehicle and a tagged RSU, the signal power received by a tagged receive RSU will experience larger path loss, thereby reducing coverage performance; when the vehicle transmission power increases, the interference transmission power of the RSU remains unchanged in this case, while the useful signal power of the tagged receiving node will increase, thereby achieving better coverage performance. Fig.~\ref{fig8} shows the expected data rate under different vehicle transmission power and different distances between the tagged transmitter-receiver. The $T$ is set to $-10$ dB. With the increase of vehicle transmission power, the expected data rate increases. When the intensity of transmit vehicles in the network is fixed, by increasing the transmission power of the transmit vehicles, the transmission rate will increase monotonically, contributing to reducing the average AoI and improving information freshness of data in the transmission process.
\begin{figure}[htbp]
	\vspace{-0.3cm}
	\setlength{\abovecaptionskip}{0.cm}
	\centering
	\includegraphics[width=0.45\textwidth]{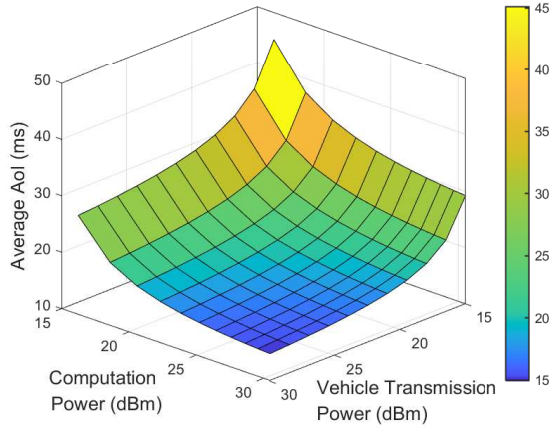}
	\caption{The trend of the average AoI versus vehicle transmission power and computation power in mesh.}
	\vspace{-0.3cm}
	\label{fig9}
\end{figure}
\begin{figure}[htbp]
	\vspace{-0.3cm}
	\setlength{\abovecaptionskip}{0.cm}
	\centering
	\includegraphics[width=0.45\textwidth]{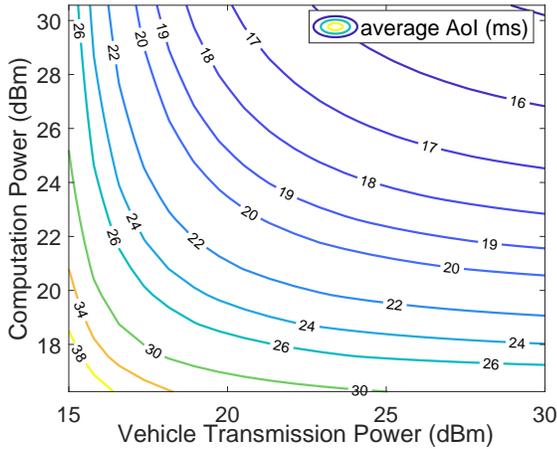}
	\caption{Communication and Computation tradeoff for average AoI.}
	\vspace{-0.3cm}
	\label{fig10}
\end{figure}

Fig.~\ref{fig9} and Fig.~\ref{fig10} show the trend of the average AoI with different vehicle transmission power and computation power. Fig.~\ref{fig9} shows that increasing the vehicle transmission power or computation power results in a better average AoI. As shown in Fig.~\ref{fig10}, when the computation capability is fixed, blindly improving the communication capacity will not effect on the obvious improvement on AoI. When computation power is $20$ dBm and average AoI is $22$ ms, the transmission power is about $25-30$ dBm. So if the transmission capacity does not match the computation capacity, it is not only difficult to guarantee the information freshness, but also a waste of resources, e.g. if the computation rate is too slow to execute tasks timely, increasing the transmission rate blindly is useless and leads to radio frequency (RF) power consumption waste and vice versa.

\section{CONCLUSION}
In this paper, the aim is to access the freshness of sensing information based on transmission-computation process in uplink VNETs scenario where vehicles are deployed for sensing data transmission and RSUs are served to enhance computation. A two-stage tandem queue is proposed to analyze the communication and computation assisted temporal performance of AoI and the Cox PPP is used to model spatial distribution of vehicles and RSUs. The closed-form solutions of the coverage probability and expected data rate as well as the analytical expressions of the average AoI of the tandem queue are derived. The simulation results verify the accuracy of these analysis and show that a tradeoff scheduling between communication and computation capacity will contribute to enhancing sensing information freshness and making efficient use of resources.

\section{ACKNOWLEDGEMENT}
This work was supported in part by the National Key R\&D Program of China under Grant (No. 2020YFB1806703), National Natural Science Foundation of China under (No. 61901044, U21A20444, 61921003, 61831002), and Young Elite Scientist Sponsorship Program by China Institute of Communications.

\section*{Appendix}
\subsection{Proof of average AoI when $\mu _1=\mu _2$}
\label{Appendix}
When $\mu _1=\mu _2$, the statistical independence characteristics based on cohort are:
\begin{equation}
	\begin{array}{l}
		f_T(t)=\left( \mu -\lambda \right) ^2te^{-\left( \mu -\lambda \right) t}.\\
	\end{array}
\end{equation}

Then we can derive the conditional expectation of $\omega _{n}^{1}$ when $Y_n=y$ and $\mu _1=\mu _2$:
\begin{equation}
	\begin{array}{l}
		E\left[ \omega _{n}^{1}|Y_n=y \right]
		=\int_y^{+\infty}{\left( t-y \right)}\left( \mu -\lambda \right) ^2te^{-\left( \mu -\lambda \right) t}dt\vspace{1ex} \\
		=ye^{-\left( \mu -\lambda \right) y}+\frac{2e^{-\left( \mu -\lambda \right) y}}{\mu -\lambda}.\\
	\end{array}
\end{equation}

Next the expectation of $\omega _{n}^{1}Y_n$ is given as:
\begin{equation}
	\begin{array}{l}
		E\left[ \omega _{n}^{1}Y_n \right] =\int_0^{\infty}{y\cdot E\left[ \omega _{n}^{1}|Y_n=y \right] \cdot f_Y\left( y \right) dy}\vspace{1ex} \\
		=\int_0^{\infty}{y\cdot \left[ ye^{-\left( \mu -\lambda \right) y}+\frac{2e^{-\left( \mu -\lambda \right) y}}{\mu -\lambda} \right] \cdot \lambda e^{-\lambda y}dy}\\
		=\lambda \left[ \frac{2}{\mu ^3}+\frac{2}{\left( \mu -\lambda \right) \mu ^2} \right].\\
	\end{array}
\end{equation}


so the average AoI when $\mu _1=\mu _2$ is as follows: 
\begin{equation}
	\begin{array}{l}
		avgAoI=\frac{E\left[ \omega _{n}^{1}Y_n \right] +E\left[ \tau _nY_n \right] +E\left[ \frac{Y_{n}^{2}}{2} \right]}{E\left[ Y_{n} \right]}\vspace{1ex} \\
		\approx \frac{\lambda \left[ \frac{2}{\mu ^3}+\frac{2}{\left( \mu -\lambda \right) \mu ^2} \right] +\frac{1}{\lambda}\cdot \left( \frac{1}{\mu}+\frac{1}{\mu} \right) +\frac{1}{\lambda ^2}}{\frac{1}{\lambda}}
		=\frac{2\lambda ^2}{\mu ^3}+\frac{2\lambda ^2}{\left( \mu -\lambda \right) \mu ^2}+\frac{2}{\mu}+\frac{1}{\lambda},\\
	\end{array}
\end{equation}
the proof is finished.

\subsection{Proof of $P\left( P_v,T \right)$}
In the uplink VNETs scenario, the coverage probability can be calculated as
\begin{equation}
\begin{array}{l}
P\left( P_v,T \right) =P\left[ \gamma _r\geqslant T \right]
=P\left[ h_r\geqslant \frac{T\left\| X_r \right\| ^{\alpha}}{P_v}\left( I_{\phi_{l_0}}+I_{\varPhi /{\phi_{l_0}}} \right) \right] 
\\
\overset{\left( a \right)}{=}\sum_{i=1}^{N_K}{w_{i}^{K}\mathcal{L} _{I_{\phi_{l_0}}}\left( \omega \right) \mathcal{L} _{I_{\varPhi /{\phi_{l_0}}}}\left( \omega \right)},
\end{array}
\end{equation}
where (a) follows that the interference can be approximated by (2) for Rician fading and set $\omega =\frac{u_{i}^{K}T\left. \left\| X_r \right\| \right. ^{\alpha}}{P_v}$. 

The Laplace transform of the interference of RSUs from the same road can be derived as
\begin{equation}
	\begin{array}{l} 
	E\left[ \prod_{\left\| X_{l}^{r} \right\| \in {\phi_{l_0}}}{e^{-\omega P_rg_l\left\| X_{l}^{r} \right\| ^{-\alpha}}} \right] 
	\\
	\overset{\left( a \right)}{=}\exp \left( -\int\limits_0^{\infty}{E_{g_l}\left[ 1-e^{-\omega P_rg_l\left\| X_{l}^{r} \right\| ^{-\alpha}} \right] \lambda _{pr}\left\| X_{l}^{r} \right\| d\left\| X_{l}^{r} \right\|} \right) 
	\\
	\overset{\left( b \right)}{=}\exp \left( -2\lambda _{pr}\left( wP_r \right) ^{\frac{1}{\alpha}}E\left[ g_{l}^{\frac{1}{\alpha}} \right] \varGamma \left( 1-\frac{1}{\alpha} \right) \right) ,
	\end{array} 
\end{equation}
where (a) follows that expectation of the independent PPP can act on Rayleigh fading $g_l\thicksim\exp\left(1\right) $, and (b) follows the properties of the Gamma function $E\left[ g_{l}^{\frac{1}{\alpha}} \right] \varGamma \left( 1-\frac{1}{\alpha} \right) =\varGamma \left( 1+\frac{1}{\alpha} \right) \varGamma \left( 1-\frac{1}{\alpha} \right) =\frac{\pi}{\sin \left( \frac{\pi}{\alpha} \right) \alpha}$. For the interference from vehicles, the close-form solution can be obtained by a similar method. Then the interference of nodes on the same road can be derived by substituting (16) and the solution to $\mathcal{L} _{I_{{\phi_{l_0}}}}\left( \omega \right) =E\left[ \prod_{\left\| X_{l}^{r} \right\| ,\left\| X_{l}^{v} \right\| ^{-\alpha}\in {\phi_{l_0}}}{e^{-\omega P_rg_l\left\| X_{l}^{r} \right\| ^{-\alpha}-\omega P_vg_l\left\| X_{l}^{v} \right\| ^{-\alpha}}} \right] $
\begin{equation}
	{\setlength\abovedisplayshortskip{1pt}}
	{\setlength\belowdisplayshortskip{1pt}}
	\begin{array}{l}
\mathcal{L} _{I_{\phi_{l_0}}}\left( \omega \right) =\exp \left( -\frac{2\pi\left( u_{i}^{K}T \right) ^{\frac{1}{\alpha}}\left. \left\| X_r \right\| \right. }{\sin \left( \frac{\pi}{\alpha} \right) \alpha}\left( \lambda _{pr}\left( \frac{P_r}{P_v} \right) ^{\frac{1}{\alpha}}+\lambda _{pv} \right) \right). 
	\end{array} 
\end{equation}

Next is the Laplace transform of the interference of RSUs from different roads
\begin{equation}
	\begin{array}{l}
		E\left[ \prod_{\left\| X_{j}^{r} \right\| \in \varPhi /{\phi_{l_0}}}{e^{-\omega P_rg_l\left\| X_{j}^{r} \right\| ^{-\alpha}}} \right] 
		\\
		\overset{\left( a \right)}{=}\prod_{\rho _j\in \varPsi}{\exp \left( -2\lambda _{pr}\int_0^{\infty}{\frac{wP_r\left( \sqrt{{\rho _j}^2+t^2} \right) ^{-\alpha}}{1+wP_r\left( \sqrt{{\rho _j}^2+t^2} \right) ^{-\alpha}}dt} \right)}
		\\
		\overset{\left( b \right)}{=}\exp \left( 2\pi \lambda _l\int_0^{\infty}{1-e^{-2\lambda _{pr}\varDelta \left( v \right)}dt} \right)
		\\
		\overset{\left( c \right)}{\approx}\exp \left( 2\pi \lambda _l\int_0^{\infty}{2\lambda _{pr}\varDelta \left( v \right) dv} \right) 
		\\
		\overset{\left( d \right)}{=}\exp \left( -\pi ^2\lambda _l\lambda _{pr}\left( wP_r \right) ^{\frac{2}{\alpha}}\frac{2\pi}{\sin \left( \frac{2\pi}{\alpha} \right) \alpha} \right), 
	\end{array} 
\end{equation}
where (a) follows the Euclidean distance from the transmitter to the tagged RSU and the probability generating function (PGFL) of the 1-D PPP $l\left( \rho _j,\theta _j \right) $, and (b) follows $\varDelta \left( v \right) =\int_0^{\infty}{\frac{wP_r\left( \sqrt{v^2+t^2} \right) ^{-\alpha}}{1+wP_r\left( \sqrt{v^2+t^2} \right) ^{-\alpha}}dt}$ and PGFL of independent line process $\varPsi $\cite{12}, and (c) using Taylor expansion to obtain the first two components with a large proportion for computation approximation:  $\mathrm{e}^x=\sum\nolimits_{s=0}^{+\infty}{\frac{x^s}{s!}}=1+x+o\left( x \right) $.
$2\lambda _{pr}\Delta \left( v \right)$ is small enough to be ignored when $s>1$. Step (d) uses the transformation of the polar coordinate.

The close-form solution can also be obtained for the interference from vehicles using the similar method as (18). Then the interference of nodes on different roads can be derived by substituting (18) and the solution to $
\mathcal{L} _{I_{\varPhi /{\phi_{l_0}}}}\left( \omega \right) =E\left[ \prod_{\left\| X_{j}^{r} \right\| ,\left\| X_{j}^{v} \right\| ^{-\alpha}\in \varPhi /{\phi_{l_0}}}{e^{-\omega P_rg_l\left\| X_{j}^{r} \right\| ^{-\alpha}-\omega P_vg_l\left\| X_{j}^{v} \right\| ^{-\alpha}}} \right] $
\begin{equation}	
	\begin{array}{l}
\mathcal{L} _{I_{\varPhi /{\phi_{l_0}}}}\left( \omega \right) \\=\exp \left( -\frac{2\pi ^3\lambda _l\left( u_{i}^{K}T \right) ^{\frac{2}{\alpha}}\left\| X_r \right\| ^2}{\sin \left( \frac{2\pi}{\alpha} \right) \alpha}\left( \lambda _{pr}\left( \frac{P_r}{P_v} \right) ^{\frac{2}{\alpha}}+\lambda _{pv} \right) \right).
	\end{array}
\end{equation}

Substituting (17) and (19) into (15) yields the result.

\end{document}